\documentclass[12pt]{article}
\usepackage{newtxtext,newtxmath}
\usepackage{graphicx}
\usepackage[letterpaper,margin=1in]{geometry}
\renewenvironment{abstract}
	{\quotation}
	{\endquotation}
\date{}

\makeatletter
\renewcommand{\fnum@figure}{\textbf{Fig.\ \thefigure}}
\renewcommand{\fnum@table}{\textbf{Table \thetable}}
\makeatother
\usepackage{url}
\newcommand{\mycomment}[1]{}
\newcommand{\email}[1]{\href{mailto:#1}{#1}}

\newcommand{\openone}{\leavevmode\hbox{\small1\normalsize\kern-.33em1}}

\usepackage{titling}
\usepackage[]{authblk}
\let\affiliation\affil

\usepackage[
 nomarkers,
 nolists,
 disable, 
]{endfloat}

\usepackage{standalone}
\usepackage{subfiles}

\usepackage{graphicx}
\usepackage{dcolumn}
\usepackage{bm}
\usepackage{physics}
\usepackage{lipsum}
\usepackage{color}
\usepackage{setspace}
\usepackage[hyperfootnotes=false,breaklinks=true]{hyperref}
\usepackage{cite}

\makeatletter
\renewcommand\@biblabel[1]{#1.}
\makeatother
\hypersetup{
 bookmarksopen=true,
 bookmarksopenlevel=1,
 colorlinks=true,
 linkcolor=blue,
 anchorcolor=blue,
 citecolor=blue,
 filecolor=blue,
 urlcolor=blue,
 pdfpagemode=UseOutlines,
 pdfstartview={XYZ null null 1},
}

\usepackage{xspace}
\newcommand{\vbm}{\ensuremath{\mathrm{V}_\mathrm{B}^-}\xspace}
\newcommand{\vbo}{\ensuremath{\mathrm{V}_\mathrm{B}^0}\xspace}
\newcommand{\Nft}{\ensuremath{\mathrm{h}^{10}\mathrm{B}^{15}\mathrm{N}}\xspace}
\usepackage{orcidlink}

\usepackage[
 draft, 
 commentmarkup=uwave,
 authormarkup=superscript, 
 authormarkupposition=left,
]{changes}
\newcommand{\addAuthor}[2]{\definechangesauthor[name={#1}, color=#2]{#1}}

\newcommand{\AuthorColorList}{%
  Alex/green,%
  Yan/cyan,%
  Huan/orange,%
  Peter/purple}

\addAuthor{Alex}{green}
\addAuthor{Yan}{cyan}
\addAuthor{Huan}{orange}
\usepackage{pgffor}
\foreach \id/\clr in \AuthorColorList
{%
 \expandafter\xdef\csname \id Add\endcsname##1{\noexpand\added[id=\id]{##1}}
 \expandafter\xdef\csname \id Delete\endcsname##1{\noexpand\deleted[id=\id]{##1}}
 \expandafter\xdef\csname \id Comment\endcsname##1{\noexpand\comment[id=\id]{##1}}
 \expandafter\xdef\csname \id Replace\endcsname##1##2{\noexpand\replaced[id=\id]{##2}{ ##1}}
}

\def\scititle{Probing Boron Vacancy Defects in hBN via Single Spin Relaxometry}
\title{\bfseries \boldmath \scititle}%

\author{Alex L.~Melendez\,\orcidlink{0009-0003-1610-1340}}
\affiliation[1]{%
 Center for Nanophase Materials Sciences, Oak Ridge National Laboratory, Oak Ridge, TN 37831, USA.
}%

\author{Ruotian Gong\,\orcidlink{0009-0006-6167-4326}}
\affiliation[2]{%
 Department of Physics, Washington University in St.~Louis, St.~Louis, MO 63130, USA.
}%

\author[2]{Guanghui He\,\orcidlink{0009-0009-7307-7694}}

\author{Yan Wang\,\orcidlink{0000-0002-6545-6434}}
\affiliation[3]{%
 Computational Sciences and Engineering Division, Oak Ridge National Laboratory, Oak Ridge, TN 37831, USA.
}%

\author{Yueh-Chun Wu\,\orcidlink{0000-0002-9299-6882}}
\affiliation[4]{%
 Materials Science and Technology Division, Oak Ridge National Laboratory, Oak Ridge, TN 37831, USA.
}%

\author{Thomas Poirier\,\orcidlink{0009-0003-5258-9160}}
\affiliation[5]{%
 Tim Taylor Department of Chemical Engineering, Kansas State University, Manhattan, KS 66506, USA.
}%

\author[1]{Steven Randolph\,\orcidlink{0000-0001-9707-4337}}

\author[1]{Sujoy Ghosh\,\orcidlink{0000-0003-2894-8971}}

\author[1]{Liangbo Liang\,\orcidlink{0000-0003-1199-0049}}

\author[1]{Stephen Jesse\,\orcidlink{0000-0002-1168-8483}}

\author[1]{An-Ping Li\,\orcidlink{0000-0003-4400-7493}}%

\author{Joshua T.~Damron\,\orcidlink{0000-0003-3409-0190}}
\affiliation[6]{%
 Chemical Sciences Division, Oak Ridge National Laboratory, Oak Ridge, TN 37831, USA.
}%

\author[4]{Benjamin J.~Lawrie\,\orcidlink{0000-0003-1431-066X}}

\author[5]{James H.~Edgar\,\orcidlink{0000-0003-0918-5964}}

\author[1]{Ivan V.~Vlassiouk\,\orcidlink{0000-0002-5494-0386}}

\author[2]{Chong Zu\,\orcidlink{0000-0001-7803-1315}}

\author[1{}*]{Huan Zhao\,\orcidlink{0000-0002-4982-0865}}
\affiliation[*]{%
Corresponding author. Email: \email{zhaoh1@ornl.gov}
}

\newcommand{\abstracttext}{%
\begin{abstract}\bfseries \boldmath

Spin defects in solids offer promising platforms for quantum sensing and memory due to their long coherence times and optical addressability. 
Here, we integrate a single nitrogen-vacancy (NV) center in diamond with scanning probe microscopy to detect, read out, and spatially map spin-based quantum sensors at the nanoscale. 
Using the boron vacancy (V$_\mathrm{B}^-$) center in hexagonal boron nitride---an emerging two-dimensional spin system---as a model, we detect its electron spin resonance indirectly via changes in the spin relaxation time ($T_1$) of a nearby NV center, eliminating the need for optical excitation or fluorescence detection of the V$_\mathrm{B}^-$. 
Cross-relaxation between NV and V$_\mathrm{B}^-$ ensembles significantly reduces NV $T_1$, enabling quantitative nanoscale mapping of defect densities beyond the optical diffraction limit and clear resolution of hyperfine splitting in isotopically enriched \Nft. 
Our method demonstrates interactions between spin sensors in 3D and 2D materials, establishing NV centers as versatile probes for characterizing otherwise inaccessible spin defects.

\end{abstract}%
}

\begin{document}
\maketitle
\abstracttext

\section*{\label{sec:Introduction}Introduction}


\quad\quad Spin-based quantum sensors leverage the coherent quantum states of localized electron or nuclear spins to detect minute variations in magnetic \cite{
    Chen_Sun_2023,
    Alsid_Danielle_2023,
    Segawa_2023}, 
electric
\cite{
    Qiu_Yacoby_2022,
    Bian_Jiang_2021}, 
thermal
\cite{
    Tzeng_Chang_2015,
    Fujiwara_2020,
    Liu_Li_2023,
    Gottscholl_Dyakonov_2021}, 
and strain fields
\cite{Lyu_Gao_2022}
with remarkable sensitivity and nanoscale resolution 
\cite{
    Degen_Cappellaro_2017,
    Wu_Lawrie_2025}.
Among the platforms developed to date, the nitrogen-vacancy (NV) center in diamond has emerged as a leading contender due to its long spin coherence times, room-temperature operation, and optical addressability  
\cite{
    Jelezko_Wrachtrup_2006,
    Schirhagl_Degen_2014,
    Du_Wrachtrup_2024}.
The spin states of NV centers can be coherently manipulated using microwave fields and optically read out, enabling a wide range of applications including nanoscale magnetometry, quantum information processing, and biological imaging 
\cite{
    Epstein_Awschalom_2005,
    Yao_Lukin_2012,
    Doherty_2013,
    Maletinsky_Jacques_2014,
    Yacoby_Casola_vanderSar_2018,
    Aslam_Park_2023}.
Despite its advantages, the NV center in bulk diamond faces significant limitations. 
In conventional architectures, NV centers typically reside tens of nanometers beneath the diamond surface, inherently limiting their proximity to external sensing targets---a critical factor for achieving ultrahigh spatial resolution. 
Moreover, diamond’s high refractive index results in significant total internal reflection, severely limiting optical fluorescence collection efficiency. 
These challenges have spurred the development of spin-based quantum sensors in atomically thin materials, wherein quantum defects reside directly at the surface, potentially enabling sub-nanometer proximity to the sensing environment and efficient photon extraction 
\cite{
    Liu_Zu_2025,
    Chen_Yan_2025}.


Currently, hexagonal boron nitride (hBN) is the only known two-dimensional material hosting optically active spin defects, making it uniquely promising for surface-proximal quantum sensing 
\cite{
    Fang_Sun_2024,
    Guo_Gu_2024
    }. 
In addition to hBN with naturally occurring isotope ratios (hBN$_\text{nat}$), isotopically purified \Nft allows clear observation of the hyperfine structure, giving insight into electron-nuclear interactions as well as an increase in magnetic field sensitivity 
\cite{
    Clua-Provost_Jacques_2023,
    Gong_Zu_2024}.
More broadly, the discovery and characterization of new spin-active defects in low-dimensional materials---some of which may ultimately emit in technologically relevant spectral ranges such as the near-infrared or telecom---requires experimental tools that do not rely on defect-specific optical readout.
A key outstanding challenge is therefore the ability to directly image the spatial distribution, density, and charge state of spin-active defects at the nanoscale. 
Conventional optical and structural probes are typically diffraction-limited, thickness-averaged, or unable to distinguish spin-active from spin-inactive charge states, masking nanoscale inhomogeneities that govern decoherence, spin–spin interactions, and device functionality. 
These limitations motivate the development of a nanoscale, charge-state-selective spin-density imaging technique that is independent of the optical properties of the target spin.
%


In this work, we use the boron vacancy (\vbm) center in hBN$_\text{nat}$ and \Nft as a model system to demonstrate that NV-based $T_1$ relaxometry can read out and spatially map arbitrary spin-active quantum defects. 
The \vbm center is a point defect consisting of a missing boron atom in the hBN lattice, exhibiting a spin-1 ground state with optically detected magnetic resonance (ODMR) at room temperature
\cite{
    Gottscholl_Dyakonov_2020,
    Gottscholl_Dyakonov_2021,
    Clua-Provost_Jacques_2024,
    Gong_Zu_2024}.
By sweeping the magnetic field through the NV-\vbm spin cross-relaxation (CR) condition and monitoring the resulting changes in the NV center’s spin relaxation time, we indirectly detect the boron vacancy’s electron spin resonance (ESR) spectrum in hBN$_\text{nat}$---eliminating the need for direct optical excitation, microwave driving, or emission-based readout of \vbm defects. 
In addition, we demonstrate the ability of cross-relaxation spectroscopy to resolve the hyperfine structure in \Nft, revealing nuclear spin information. 
Finally, we perform nanoscale spatial mapping of the cross-relaxation signal, revealing spatial heterogeneity in spin noise and quantitatively correlating this noise with V$_\mathrm{B}^-$ defect densities. 
Critically, our method addresses a key limitation of conventional defect-mapping techniques (e.g., Raman, electron microscopy) that cannot distinguish neutral from negatively charged boron vacancies, demonstrating that fewer than 10\% of the vacancies are negatively charged and thus suitable as quantum sensors. 
These findings emphasize the versatility of scanning NV cross-relaxometry as a powerful method for probing emerging quantum spin systems, especially those with limited optical accessibility.

\section*{\label{sec:Results}Results}

\begin{figure}
\centering
\includegraphics[width=0.99\textwidth]{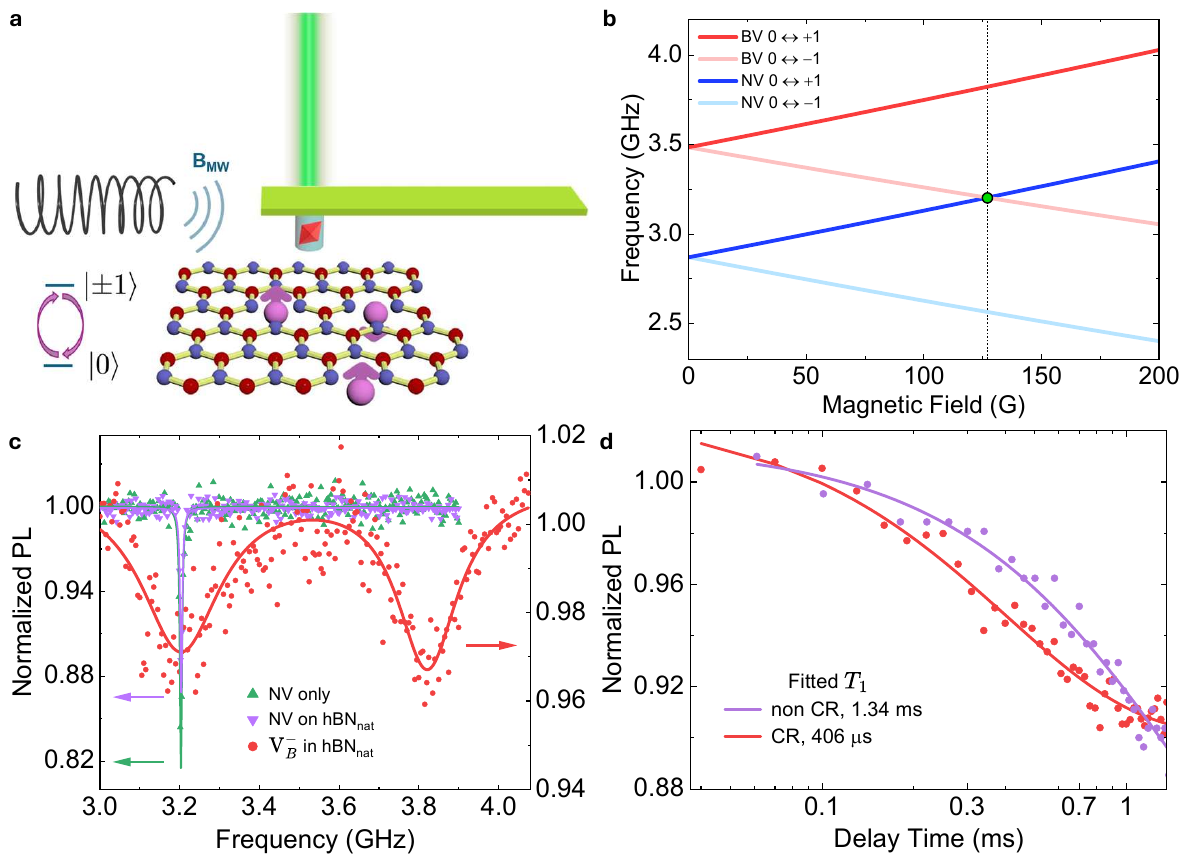}
\caption{\label{fig:1} 
{\setstretch{1.0}\small{\textbf{NV $T_1$ relaxometry at CR condition with \vbm in hBN$_\text{nat}$:}
(a) A scanning NV microscope containing a single NV center is brought near a 90-300\,nm thick hBN$_\text{nat}$ sample that has been irradiated with He ions to create an ensemble of \vbm centers. 
The NV center's photoluminescence (PL) is collected using a confocal optical setup.
A microwave field $B_\text{MW}$ is supplied to both the NV and the nearest \vbm centers by an antenna positioned near the cantilever tip.
When the microwave frequency matches the spin resonance frequency of the NV or \vbm centers, transitions between the ground state sublevels $m_s=0\leftrightarrow\pm1$ states are driven coherently.
(b) Calculated spin resonance dispersions of the NV and \vbm centers with the magnetic field oriented 29\textdegree\ from the surface normal. 
Cross-relaxation occurs near 127\,G bias, where the NV $m_s = 0 \leftrightarrow +1$ transition overlaps with the \vbm $m_s = 0 \leftrightarrow -1$ transition, highlighted by the green dot.
(c) CW-ODMR spectra of the NV center before (green curve) and after (purple curve) engaging with the hBN sample, and of the \vbm centers in hBN (red curve), measured under the cross-relaxation condition shown in (b). 
All spectra are fitted using Lorentzian functions. The left and right vertical axes correspond to the NV and \vbm data, respectively. Only the $m_s = 0 \leftrightarrow +1$ transition of the NV is shown.
The arrows denote via their colors which vertical axis each trace is plotted against.
(d) NV spin $T_1$ measurement in non-CR condition (purple curve, fitted $T_1 = 1.34 \pm 0.19$\,ms) and CR condition (red curve, $T_1 = 406 \pm 34\,\mu$s), both measured after engaging the hBN$_\text{nat}$ sample. Curves are single exponential fits $
\propto\exp(-t/T_1)$.
}}
}
\end{figure}

We investigated both chemical vapor deposition (CVD)-grown natural-abundance hBN (hBN$_\text{nat}$) and mechanically exfoliated isotopically enriched \Nft (99\% $^{10}$B, 99.5\% $^{15}$N). 
The hBN$_\text{nat}$ membrane, with a nonuniform thickness ranging from 90 to 300\,nm, was irradiated with helium ions to generate boron vacancy defects at an estimated density of $\sim$0.1 defects per nm$^2$ per atomic layer (5400 ppm) 
\cite{
    Vlassiouk_2025,
    Sarkar_Gradecak_2024}. 
Isotopically purified \Nft crystals were synthesized via the atmospheric pressure high-temperature (APHT) method, neutron-irradiated, and subsequently exfoliated into flakes with an average thickness of $\sim$50\,nm. 
A single NV center, oriented along the diamond (100) plane and located approximately $9 \pm 4$\,nm beneath the surface, was integrated into the apex of a tuning fork-based scanning probe cantilever to form a scanning NV magnetometer. 
Operating in frequency-modulated contact mode with a finely tuned frequency offset, the best NV-to-sample distance was calibrated to be approximately $11.4\pm1.5$\,nm (see Supplementary Note S1).
Optical excitation and fluorescence collection from the NV center were achieved using a confocal microscope, while microwave fields for spin manipulation of both the NV and \vbm centers were delivered through an integrated radio-frequency (RF) antenna (Fig.~\ref{fig:1}a).

Figure \ref{fig:1}b presents calculated ESR spectra for both the NV and \vbm centers under a magnetic field applied at an angle of 29\textdegree\ relative to the NV axis. 
At a bias field of 127\,G, the NV $m_s=0\leftrightarrow+1$ and \vbm $m_s=0\leftrightarrow-1$ spin transitions become energetically degenerate at approximately 3.2\,GHz. 
Under this resonance condition, the NV and \vbm spins can exchange energy non-radiatively via magnetic dipole–dipole coupling---a process known as cross-relaxation (CR) \cite{Bajaj_2014,Wood_Hollenberg_2016}---which results in an enhanced relaxation rate of the NV spin. 
Figure \ref{fig:1}c shows continuous-wave ODMR (CW-ODMR) measurements of the NV center (only the $m_s=0\leftrightarrow+1$ transition is plotted) and the \vbm ensemble in hBN$_\text{nat}$ under the CR condition calculated in \ref{fig:1}b, confirming spectral overlap of their respective spin transitions. 
Compared to the single NV center, the \vbm ensemble ODMR spectrum exhibits significantly broader linewidths and lower contrast, limiting its performance as a standalone quantum sensor.
Upon bringing the NV sensor into contact with the hBN sample, the ODMR contrast of the NV $m_s = 0 \leftrightarrow +1$ transition decreases from 18.5\% to 13.1\%. 
This reduction in contrast may reflect faster spin population equalization, i.e., a shorter $T_1$, though other processes such as charge dynamics may also explain this decrease 
\cite{yuan2020charge}.
The $T_1$ reduction is confirmed by full $T_1$ measurements, as shown in Figure~\ref{fig:1}d. 
While engaged with the sample, a magnetic field is applied to bring the NV and \vbm into the CR condition, decreasing the NV spin $T_1$ from $1.34 \pm 0.19$\,ms to $406 \pm 34\,\mu$s.


The total relaxation rate $\Gamma_1=1/T_1$ of the NV center can here be modeled as a contribution from three terms \cite{Gong_2026_review}:
\begin{align}
    \Gamma_1 = \Gamma_1^\text{int} + \Gamma_1^\text{bath} + \Gamma_1^\text{CR}, 
\end{align}
where $\Gamma_1^\text{int}$ is the intrinsic relaxation rate of the NV, $\Gamma_1^\text{bath}$ is the contribution from the ambient spin noise produced by the hBN, and $\Gamma_1^\text{CR}$ is the contribution due to the CR condition being met between the NV and \vbm spins. 
Taking dipole-dipole interaction between the \vbm spin bath and the NV as the dominant interaction, the ensemble-averaged relaxation rate is \cite{He_Zu_2023,Scholten2024}
\begin{align}\label{CR_Rate}
    \Gamma_{1}^\text{CR} 
 &= \expval{\left(\frac{J_0\mathcal{A}_j}{r_j^3}\right)^2
    \frac{2}{\gamma+\Gamma_2^\text{NV}+\Gamma_2^{\vbm}}},
\end{align}
where $J_0=2\pi\times52\,\text{MHz}{\cdot}\text{nm}^3$ is the dipolar coupling strength, $\mathcal{A}_j$ contains the angular dependence between the NV and $j^\text{th}$ \vbm spin, $\Gamma_2^\text{NV}$ and $\Gamma_2^{\vbm}$ are the NV and \vbm transition linewidths, and $\gamma$ is the interaction-induced intrinsic linewidth. 
Using Eq.~\eqref{CR_Rate}, a Monte Carlo simulation of the NV-\vbm dipolar interactions was performed showing that the cross-relaxation rate of $\Gamma_1^\text{CR}=1.72\,$kHz seen in Fig.\,\ref{fig:1}d corresponds to a \vbm density of about 220\,ppm (see Supplementary Note S4 for details).

\begin{figure}
\centering
\includegraphics[width=0.99\textwidth]{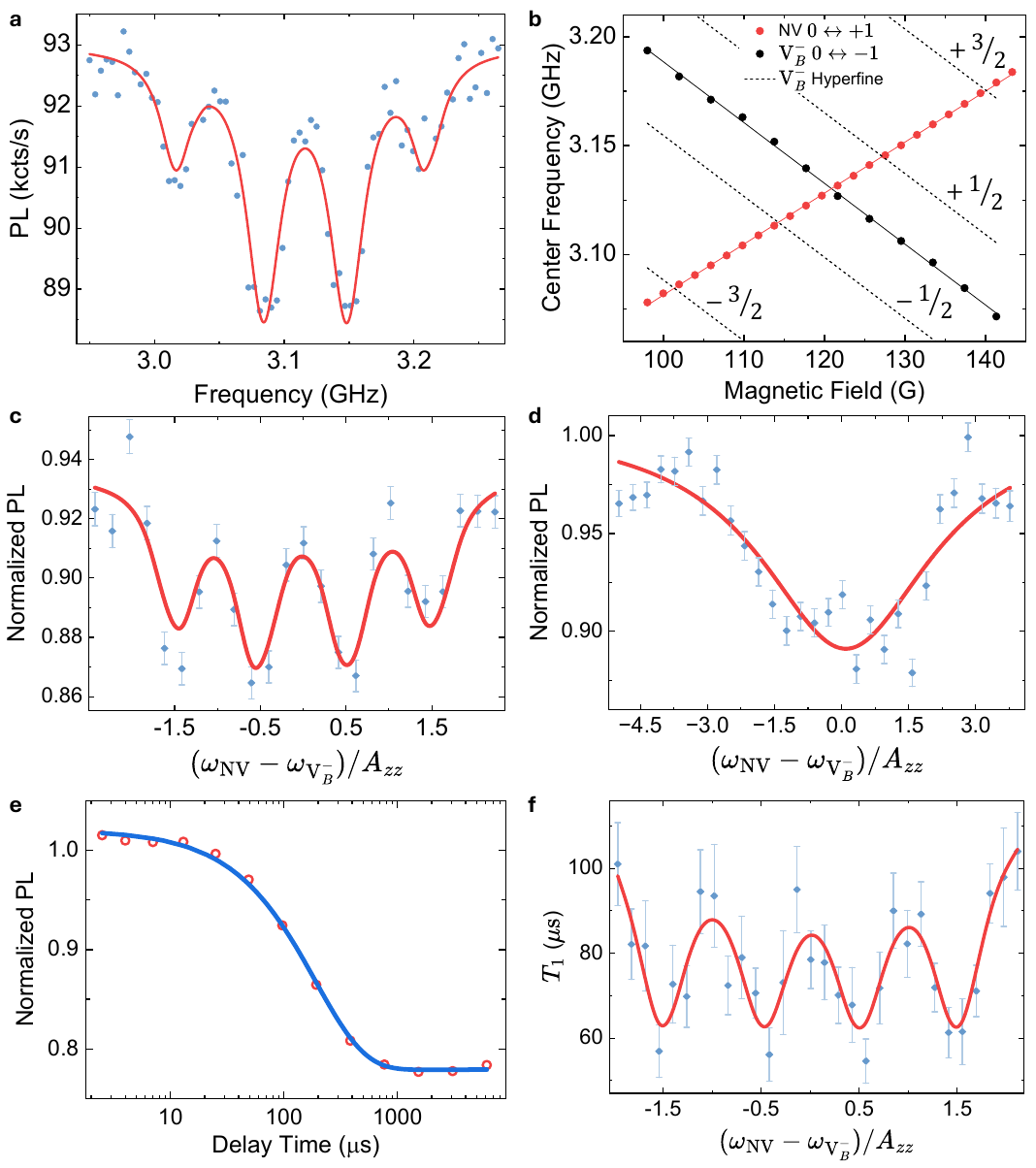}
\end{figure}

\begin{figure}
\centering
\caption{\label{fig:2} 
{\setstretch{1.0}\small{\textbf{$T_1$-MR detection of hyperfine structure:}
(a) CW-ODMR spectrum of \vbm centers in \Nft showing four hyperfine-split lines, fitted with Lorentzian functions (red). 
A magnetic field of 123\,G is applied out-of-plane. 
(b) Extracted ODMR center frequencies of the NV center (red) and \vbm centers (black) as a function of magnetic field, with solid lines representing fitted curves. 
The hyperfine line positions (dashed) are calculated using the average $A_{zz}$, revealing four expected cross-relaxation conditions corresponding to the crossover points between the NV and \vbm transitions. 
(c) NV single-$\tau$ $T_1$-MR detection of \vbm hyperfine structure by sweeping the magnetic field through the four CR conditions. 
The magnetic field is converted to frequency detuning between the NV and \vbm transitions and plotted in units of $A_{zz}$ in \Nft (67\,MHz).
The data are fitted using a sum of four Lorentzian functions, where dips 1 and 3 share one set of parameters, and dips 2 and 4 share another (fit shown in red). 
Here, the free evolution time $\tau$ is chosen to be 2\,ms, given that the tip's non-CR $T_1$ measured immediately before this $T_1$-MR measurement is $2.09 \pm 0.25$\,ms (see Supplementary Figure S23a). 
(d) NV single-$\tau$ $T_1$-MR measurement of hBN$_\text{nat}$ under varying magnetic field. 
The field axis is converted to frequency detuning and plotted in units of $A_{zz}$ in hBN$_\text{nat}$ (44\,MHz).
Data are fitted using a Lorentzian function (red curve). The $\tau$ in this case is 0.7\,ms, and the tip's $T_1$ measured at the non-CR condition is $1.21 \pm 0.32$\,ms (see Supplementary Figure S23b).
Error bars of (c) and (d) represent normalized shot noise level.
(e) A short-$T_1$ NV tip's spin $T_1$ relaxation curve before engaging the sample, showing $T_1$ of $191 \pm 7$\,$\mu$s.
(f) NV $T_1$ values measured as a function of magnetic field (blue symbols with error bars), showing four distinct cross-relaxation dips. The curve is fitted with a sum of Lorentzian functions and plotted in units of $A_{zz}$ in \Nft (67\,MHz). Error bars represent $1\sigma$ uncertainties of fitted $T_1$ values.
}
}}
\end{figure}

Hyperfine splitting in ODMR spectra arises from the magnetic interaction between an electron spin and nearby nuclear spins, leading to multiple resonance lines that reflect the local nuclear spin environment. 
These splittings are essential for nuclear-spin–based sensing, as they enable identification, control, and readout of individual nuclear species, allowing for nanoscale magnetic resonance and enhanced spectral selectivity in quantum sensing applications. 
Surrounding the \vbm electron spin are three nearest-neighbor nitrogen nuclei, whose hyperfine interactions split the ODMR transitions. 
In natural-abundance hBN (hBN$_\text{nat}$), the dominant isotope is $^{14}$N ($I = 1$), which results in seven allowed hyperfine transitions. 
These lines overlap and broaden due to nuclear spin mixing and inhomogeneities, producing the broad ODMR spectra observed in Fig.~\ref{fig:1}c. 
In contrast, isotopically enriched \Nft contains $^{15}$N nuclei ($I = 1/2$), yielding four hyperfine-split lines corresponding to the total nuclear spin projection $m_I = \{-3/2, -1/2, +1/2, +3/2\}$~\cite{Clua-Provost_Jacques_2023}.

Figure~\ref{fig:2}a shows a representative CW-ODMR spectrum of the \vbm $m_s = 0 \leftrightarrow -1$ transition in \Nft, exhibiting clearly resolved hyperfine structure. 
In total, 24 CW-ODMR spectra were recorded for the NV $m_s = 0 \leftrightarrow +1$ and 12 for the \vbm $m_s = 0 \leftrightarrow -1$ transitions as a function of an out-of-plane (OOP) magnetic field, with the extracted center frequencies plotted in Fig.~\ref{fig:2}b. 
The average hyperfine splitting between the $m_I = \pm 1/2$ levels is $|A_{zz}| = 2\pi \times (67 \pm 2)$~MHz (see Supplementary Note S5 for additional data). 
Using these values, the positions of the four hyperfine transitions (dashed lines) are plotted, from which four distinct cross-relaxation points with the NV $m_s = 0 \leftrightarrow +1$ transition are identified at 101, 114, 127 and 140~G.

To rapidly assess the magnetic field dependence of NV spin relaxation, we employed a single-$\tau$ $T_1$ relaxometry technique. 
In this method, the NV center is first polarized into the $m_s = 0$ spin state, followed by a fixed evolution time $\tau$, after which the photoluminescence (PL) intensity is recorded. 
A reduced ratio $\mathrm{PL}(\tau)/\mathrm{PL}(0)$ indicates enhanced relaxation and thus a shorter $T_1$ time. 
Figure~\ref{fig:2}c presents $\mathrm{PL}(\tau)/\mathrm{PL}(0)$ as a function OOP magnetic field after engaging the NV tip with an exfoliated \Nft{} flake, revealing four dips corresponding to the hyperfine transitions of \vbm{} centers in \Nft{}. These features occur when the NV ESR frequency matches the hyperfine-split ESR frequencies of the \Nft, confirming that our microwave-free cross-relaxation method can probe both electron and nuclear spin properties and recover the spin resonance spectra.

For comparison, we performed the same field-dependent $\mathrm{PL}(\tau)/\mathrm{PL}(0)$ measurement after landing an NV tip on a hBN$_\text{nat}$ sample. 
As shown in Fig.~\ref{fig:2}d, the resulting curve closely resembles the CW-ODMR spectrum of \vbm in hBN$_\text{nat}$ given in Fig.~\ref{fig:1}c. 
Compared to the \vbm ODMR spectrum, the field-dependent single-$\tau$ $T_1$ relaxometry curve---hereafter referred to as the $T_1$-magnetic resonance ($T_1$-MR) curve---exhibits enhanced signal contrast, limited primarily by the intrinsic spin-dependent fluorescence contrast of the NV center. Additional $T_1$-MR data acquired on the same sample using an independent NV probe are presented in Supplementary Note~7, demonstrating that the measured $T_1$-MR response is reproducible across different tips.


Diamond tips with long NV $T_1$ times offer higher sensitivity but require significantly longer acquisition times to record a complete $T_1$ decay curve, making single-$\tau$ $T_1$-MR a more realistic approach for field-dependent measurements. 
However, the $\mathrm{PL}(\tau)/\mathrm{PL}(0)$ signal is also influenced by the NV readout contrast, which in our experiments decreases approximately linearly with increasing magnetic field (Supplementary Fig.~S21).
This introduces a smooth, non-resonant background in single-$\tau$ $T_1$-MR traces that is unrelated to cross-relaxation.
To remove this contribution, each $T_1$-MR curve in Fig.~\ref{fig:2} was baseline-corrected by fitting the non-resonant background to a linear function and subtracting it, thereby isolating the cross-relaxation features.
A derivation and full description of this baseline-correction procedure are provided in Supplementary Note~5.

NV tips with shorter $T_1$ times, although less sensitive, allow acquisition of full $T_1$ decay curves at each field within a reasonable timescale, thereby enabling direct extraction of $T_1$ values and eliminating the influence of field-dependent ODMR contrast. 
All prior measurements presented in this study were performed with tips having $T_1 > 1\,$ms to maximize sensitivity (see Supplementary Information Table S1 for a list of NV tips used). 
In contrast, Figure~\ref{fig:2}e shows relaxation from a tip with a significantly shorter $T_1$ of $191\,\mu$s. 
Once the tip engages with a \Nft flake, full $T_1$ relaxation curves are acquired at varying magnetic fields, and the corresponding $T_1$ values are extracted. Figure~\ref{fig:2}f displays the $T_1$-field dependence, clearly resolving the hyperfine structure of \Nft.
We note that the relative dip depths in $T_1$-MR can deviate from the hyperfine intensity ratios observed in direct ODMR, particularly for short-$T_1$ probes where broadened response and partial saturation/nonlinear cross-relaxation can equalize the apparent contrasts.
Importantly, these effects do not affect the extracted resonance frequencies or linewidths, which underpin the analysis.

\begin{figure}
\centering
\includegraphics[width=0.98\textwidth]{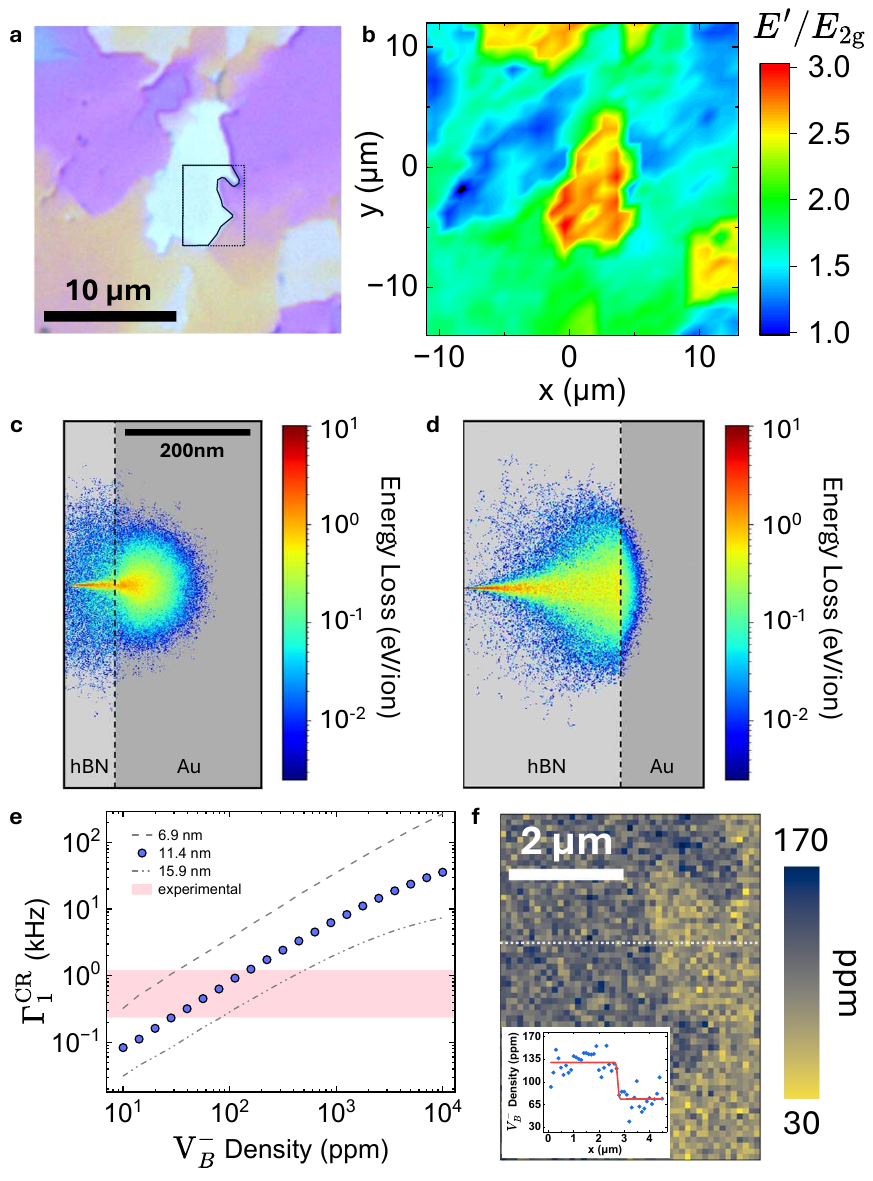}
\end{figure}

\begin{figure}
\centering
\caption{\label{fig:3} 
{\setstretch{1.0}\small{\textbf{Nanoscale imaging of \vbm density in hBN$_\text{nat}$ using NV-\vbm cross relaxometry:}
(a) Optical microscope image of hBN$_\text{nat}$ sample on gold surface, with different colors corresponding to variations in the hBN thickness. 
The light-colored area in the center has thickness $\sim$90\,nm compared to the surrounding pink/purple areas with thickness $\sim$190\,nm.
Here a portion of the right edge of the light-colored area is highlighted in a blue polygon which was measured by the scanning NV in panel (f). 
(b) Raman spectroscopy map showing the ratio of the $E'$ peak to the $E_{\text{2g}}$ peak across the sample area displayed in panel (a). 
(c) SRIM simulations of the ion energy losses (normalized per ion) of 30\,keV He$^+$ (10,000 ions simulated) for 90\,nm and (d) 250\,nm hBN on gold, suggesting thickness-dependent non-uniform defect density distribution.
(e) Monte Carlo simulation of the cross-relaxometry contribution to the NV relaxation rate $\Gamma_1^\text{CR}$ as a function of the \vbm density and three different NV-to-sample distances. 
Horizontal gray bar shows values of $\Gamma_1^\text{CR}$ associated with range of the color scale in (f).
(f) \vbm density map, obtained as a single-$\tau$ $T_1$ spatial scan over a free-evolution time $\tau=250 \,\mu$s which was converted to \vbm density via the Monte Carlo simulation in panel (e). 
The size of each pixel is 100\,nm. The measurement was conducted at the CR condition.
Inset: The profile of \vbm density as a function of position, over the line indicated by the white dashed line in (f). The data is fit to the edge-spread function (red) yielding a step width of 46\,nm. 
}}
}
\end{figure}

Understanding the spatial distribution of spin defects is essential not only for optimizing defect creation and implantation processes but also for probing nanoscale spin--spin interactions that govern decoherence and quantum entanglement. 
Raman spectroscopy, particularly through mapping of defect-associated vibrational modes, has long served as a valuable technique for visualizing defect distributions 
\cite{
    Dresselhaus_Saito_2010,
    Eckmann_Casiraghi_2012,
    Lee_Jeong_2022,
    Zhang_2024}. 
However, its diffraction-limited spatial resolution ($\sim$500\,nm) restricts access to nanoscale variations that are often critical in quantum materials. 
More importantly, most conventional defect mapping techniques---including Raman and transmission electron microscopy (TEM)---cannot distinguish between neutral (\vbo) and negatively charged (\vbm) boron-vacancy centers. 
Since only \vbm possesses the desired spin properties and optical addressability, it is crucial to develop spatial mapping methods that selectively target \vbm defects.
Using the NV center’s spin $T_1$ as imaging contrast, scanning NV microscopy offers both high spatial resolution and selective sensitivity to \vbm centers. 
When utilizing shallowly implanted NV centers, various sensing techniques have demonstrated spatial resolution down to $\sim$10\,nm, enabling direct imaging of magnetic noise from paramagnetic spin distributions with nanometric precision 
\cite{
    Rugar_Awschalom_2015,
    wang2019nanoscale}. 
Under cross-relaxation conditions, regions of the hBN sample with higher surface \vbm densities generate stronger magnetic noise at the NV resonance frequency, leading to enhanced NV spin relaxation. 
By raster-scanning the NV sensor and performing single-$\tau$ $T_1$ relaxometry at each pixel, a high-resolution spatial map of \vbm density can be constructed, revealing nanoscale variations in defect distribution well below the diffraction limit.


Figure \ref{fig:3}a shows the optical image of a hBN$_\text{nat}$ flake after helium ion irradiation, while Figure \ref{fig:3}b presents the corresponding Raman intensity map of the $E'/E_{\text{2g}}$ vibrational peaks. 
The $E_{\text{2g}}$ mode, associated with in-plane vibrations of $sp^2$-bonded boron and nitrogen atoms, is characteristic of pristine hBN
\cite{
    Krecmarova_Sanchez_2019}. 
In contrast, the $E'$ peak at $\sim$1290\,cm$^{-1}$ emerges following helium ion irradiation and is attributed to the formation of \vbm defects
\cite{
    Sarkar_Gradecak_2024,
    Linderalv_Erhart_2021}.
Stopping and Range of Ions in Matter (SRIM) simulations were performed to estimate the He$^+$ energy loss for 90\,nm vs 250\,nm thick hBN on a gold substrate, showing thickness-dependent ion backscattering (Fig.~\ref{fig:3}c-d and Supplementary Note S6).
Based on these simulations, it is expected that a higher proportion of defects is concentrated at the surface of the 90\,nm film and delocalized from the impact point. 
Conversely, it would be expected that defects in the 250\,nm film would remain localized in the near surface region, but broaden significantly throughout the thickness of the film.
Hence regions of varying hBN thickness---visible as distinct colors in the optical image---exhibit different \vbm defect densities 
\cite{Vlassiouk_2025}. 
Consequently, structural features observed in both the optical and Raman maps are spatially correlated.


To further resolve local spin defect distributions beyond the optical diffraction limit, we performed a single-$\tau$ $T_1$ relaxometry scan over the region of interest with a 100\,nm step size. 
Figure~\ref{fig:3}e shows Monte Carlo simulations of the cross-relaxation rate Eq.~\eqref{CR_Rate} at various \vbm densities, allowing conversion between a spatial map of NV relaxation to that of \vbm density depicted in Figure \ref{fig:3}f (see Supplementary Note S4).
This NV-based nanoscale mapping resolves spatial features in the defect density that are not discernible in the Raman data, including abrupt grain boundaries in the CVD-grown hBN. Additional correlative imaging of the same sample region is provided in Supplementary Note~S9.
In addition, as a quantitative measurement, it offers the spin density value at each pixel.
Notably, the spatial resolution of the scanning NV relaxometry measurement and thereby the \vbm density is primarily limited by NV-sample distance and can be improved to $\sim$10\,nm by reducing the scan step size, although this significantly increases the total acquisition time.
%


\section*{\label{sec:Discussion}Discussion}

The shot noise-limited magnetic sensitivity of a spin-1 quantum sensor operating under the CW-ODMR protocol is given by \cite{Barry_Walsworth_2020}:
\begin{align}
    \eta_{\text{CW}} \propto \frac{\Delta \nu}{C_{\text{CW}} \sqrt{R}},
\end{align}
where $\Delta\nu$ is the ODMR linewidth, $C_\text{CW}$ is the ODMR contrast, and $R$ is the photon detection rate. 
Optimal magnetic field sensitivity therefore requires a narrow ODMR linewidth, high contrast, and efficient photon collection. 
In practice, the ODMR contrast of \vbm ensembles is typically low ($<$5\%) for several reasons: 
(1) the hBN contains a fraction of bright defects that do not exhibit ODMR; 
(2) the spin-lattice relaxation time $T_1$ is typically short ($<$10\,$\mu$s), which limits spin polarization fidelity; and 
(3) the spin-dependent fluorescence contrast is intrinsically weak due to competing non-radiative decay pathways via intersystem crossing. 
In contrast, our $T_1$-MR measurement, which reproduces the ODMR spectral profile, yields higher contrast and does not rely on the optical readout of the target spin species. 
The method detects population decay from the optically polarized NV spin $m_s=0$ state toward thermal equilibrium. 
Assuming a 30\% intrinsic NV ODMR contrast, the expected change in PL from the polarized state to the thermalized population yields a theoretical PL contrast of $\sim$20\% in all-optical $T_1$ relaxometry.
When performing field-dependent full $T_1$ measurements and using the extracted $T_1$ values---rather than the $\mathrm{PL}(\tau)/\mathrm{PL}(0)$ ratio---as the sensing metric, the $T_1$-MR contrast of NV--\vbm spin resonance detection can be enhanced by an order of magnitude. 
For instance, as shown in Fig.~\ref{fig:1}d, under cross-relaxation conditions the NV $T_1$ decreases from 1.34\,ms to 406\,$\mu$s, corresponding to a contrast of approximately 70\%. 
The field-dependent full $T_1$ measurements using a short-$T_1$ NV tip, shown in Fig.~\ref{fig:2}f, reveal the hyperfine structure of \Nft with approximately 50\% $T_1$-MR contrast.


Another key advantage lies in the photon detection efficiency. 
The fluorescence spectrum of \vbm peaks at $\sim$850\,nm, whereas silicon avalanche photodiodes (APDs) are most efficient near 650\,nm, leading to sub-optimal detection. 
In contrast, our NV-based relaxometry approach relies on the NV center’s own fluorescence for readout, which allows for the detection of spin species across a broad spectral range. 
This includes optically dark defects as well as those emitting at technologically important telecom wavelengths, all using a single, low-cost APD. 
For spin defects that emit within the same spectral range as NV centers---and therefore cannot be isolated using conventional optical filters---we developed a dark readout method that enables effective separation through time-gated detection, exploiting differences in fluorescence dynamics rather than spectral properties (see Supplementary Note S2). 
We acknowledge that a limitation of relaxometry is its longer acquisition time due to its pulsed nature---typically about 3,000 times longer than that of CW-ODMR.
This corresponds to a sensitivity degradation of approximately $1/\sqrt{3000}$, or about 55 times lower sensitivity assuming equal acquisition time. 
However, this limitation can be offset by the enhancement in signal contrast and further mitigated by using dense NV ensembles, which significantly improve the signal-to-noise ratio and reduce integration time---paving the way for scalable, high-contrast, and broadband quantum sensing.


Supplementary Note~3 presents a quantitative model for single-$\tau$ $T_1$-MR measurements and derives the optimal wait time $\tau$ that maximizes signal-to-noise by balancing contrast against photon shot noise and measurement duty cycle.
Within the same framework, we evaluate the shot-noise-limited magnetic-field sensitivity $\eta(\tau)$ near the NV--\vbm cross-relaxation condition.
This analysis further indicates that $T_1$-MR readout using the bright NV fluorescence can outperform direct PL-based ODMR of \vbm centers in the few-spin (or nanoscale-cluster) limit.
In this regime, the \vbm PL from a $\sim$1--10\,nm defect cluster can be far below the level typically detected in confocal measurements that collect fluorescence from a diffraction-limited volume ($\sim$500\,nm lateral diameter).
By contrast, scanning NV cross-relaxometry can place the NV directly above the cluster, and the cross-relaxation contribution to the NV relaxation rate is dominated by the nearest defects and decays steeply with separation ($\propto r^{-6}$), enabling a measurable $T_1$-MR signal even when \vbm ODMR is not feasible.


Under helium‐ion irradiation at a dose of 50\,He$^+$/nm$^2$, scanning‐NV cross‐relaxometry reveals a \vbm density of 30--170\,ppm, which is substantially lower than the $\sim$5400\,ppm total vacancy density measured by atomic-resolution TEM imaging \cite{
    Sarkar_Gradecak_2024}. 
This indicates that most boron vacancies remain in the neutral V$_\mathrm{B}^0$ state. 
Our inferred negatively charged fraction of 3.1--9.3\% aligns with the 1--10\% \vbm density reported by a previous work \cite{Gong_Zu_2023}. 
Traditional optical methods average over regions that include both high and low defect densities, which can obscure sharp local variations. 
In contrast, our nanoscale relaxometry technique offers precise, spatially resolved measurements that preserve local contrast and reveal fine-scale features. 
In addition, unlike optical methods that integrate signals over the entire sample thickness, our technique primarily probes surface defect densities---precisely the region most critical for quantum sensing applications (see Supplementary Note 4 and Figure S17). 
A major limitation for many optical and electrical probes is that they cannot effectively distinguish \vbm from V$_\mathrm{B}^0$ or isolate the near-surface defect layer relevant for device operation. 
By tuning the magnetic field to the NV--\vbm cross-relaxation condition, our NV-based $T_1$ relaxometry becomes selectively sensitive to spin-active \vbm defects whose ESR matches that of the NV, while neutral V$_\mathrm{B}^0$ and other ESR-detuned centers contribute negligibly to the contrast. 
As a proof-of-principle relevant to device operation, we combine electrostatic gating with NV--\vbm cross-relaxometry (Supplementary Note~8) by tracking the gate-dependent change in NV $T_1^{-1}$ at the cross-relaxation field, which directly reports the surface-proximal \vbm spin density.
We observe a $\sim$30\% modulation, consistent with our surface band-bending hypothesis, in which electrostatic gating induces a band bending that alters the local defect charge-state population in a shallow, surface-proximal region of hBN (see Supplementary Note~8 for more discussion). This modulation is substantially larger than the few-percent changes inferred from thickness-averaged photoluminescence \cite{fraunie2025charge}, underscoring the charge-state selectivity and surface sensitivity of our non-invasive approach.


Previous studies have demonstrated cross-relaxation (CR) between NV centers and nearby proton spins or other spin defects within diamond, and the ability of NV centers to read out other electron spins has been established in several pioneering works 
\cite{
    Wood_Hollenberg_2017,
    Bajaj_2014,
    Wood_Hollenberg_2016,
    grinolds2013nanoscale,
    sushkov2014magnetic,
    shi2015single}. 
However, these experiments mainly rely on dynamical-decoupling protocols, which require (i) microwave driving of the NV center, (ii) microwave or RF driving of the target spin, (iii) prior knowledge of the target spin resonance (frequency and $\pi$-pulse length), and (iv) nanosecond-scale timing control; in practice, such pulsed schemes are most convenient for frequencies up to the hundreds-of-MHz range, where high-power, short $\pi$ pulses are experimentally tractable. 
Cross-relaxation between an NV and spin-based quantum sensors in external materials---especially those of different dimensionalities---has not yet been achieved, and demonstrating such interactions is critical for extending NV-based detection to a broader range of quantum systems. 
Prior CR experiments have also relied on static diamond slabs, where the spatial relationship between the NV and target spins is fixed, limiting control over interaction strength and spatial resolution. 
In contrast, our method uses NV–\vbm cross-relaxation $T_1$ relaxometry as a microwave-free, all-optical readout: the protocol (i) does not require any microwave control of the NV center, (ii) only requires microsecond–millisecond timing resolution, (iii) operates with minimal prior knowledge of the target spin, and (iv) yields an ESR spectrum at GHz frequencies set by the applied magnetic field. 
Furthermore, by combining this cross-relaxation contrast with quantitative modeling, we can not only detect the presence of target spins but also map their local spin density at the nanoscale. 
Implemented in a scanning NV geometry, this approach provides two major advantages: precise nanoscale control over the NV–target distance to optimize dipole–dipole coupling, and the capability to spatially map spin noise with sub-diffraction resolution. 
This platform also enables simultaneous manipulation and readout of spin dynamics---achieved by adjusting the NV–target separation or applying localized strain to the target using the same diamond tip employed for CR-based sensing.


In conclusion, we demonstrate NV-based spin relaxometry as a versatile method for characterizing and spatially mapping spin-active quantum defects, exemplified by boron vacancy defects in 2D hBN.  
Leveraging the established platform of scanning NV microscopy, our technique bypasses the conventional requirement for dedicated optical excitation and detection infrastructure for each new defect species. 
This enables all-optical relaxometry-based ESR measurements with high spectral contrast and nanoscale spatial precision. 
Our approach streamlines and standardizes the discovery of novel spin-active defects, including those emitting at telecom wavelengths or optically dim defects that are otherwise inaccessible with low-cost detectors.
Furthermore, our approach enables heterogeneous quantum architectures by decoupling sensing and readout functionalities into distinct qubits and thus harnessing their complementary strengths while minimizing individual limitations. 
Building upon this demonstration, we envisage future directions that include controlled gate operations between NV centers and nearby two-dimensional spin defects, entanglement-enhanced quantum sensing, and scanning-probe NMR spectroscopy via cross-relaxation protocols.


During revision of this manuscript, we became aware of a related study published by H.~Sun \textit{et al.}~\cite{sun2025room} reporting nanoscale imaging of spin defects in two-dimensional materials. Their results are complementary to those reported here.

\section*{\label{sec:Methods}Methods}

\subsection*{\label{ssec:Sample_Preparation}Sample Preparation}

Flakes of thick hBN$_\text{nat}$ were synthesized by chemical vapor deposition (CVD) using solid boron precursors in a nitrogen atmosphere. 
A flake of variable thickness (90-300\,nm) was transferred onto a gold coplanar waveguide (CPW) which was lithographically patterned onto a sapphire substrate. 
However, this CPW was not used as a waveguide for microwaves in the course of this work.
The hBN$_\text{nat}$ was irradiated with a beam of He ions at a dose of 50\,He$^+$/nm$^2$ at an energy of 30\,keV using a Zeiss Orion NanoFab Helium ion microscope. 
At this energy the penetration depth in hBN was about 240\,nm. 
As a result, more He ions reached the gold where the hBN$_\text{nat}$ was thinner, which increase the production of \vbm centers. 
%


The \Nft crystal flakes were grown by the atmospheric pressure high temperature (APHT) method
\cite{Liu_Edgar_2018,Janzen_Edgar_2024}. 
Briefly, the process starts by mixing high purity elemental boron, enriched in the $^{10}$B isotope to 99\%, with nickel and chromium with weight ratios of 4:48:48, respectively. 
The mixture is then heated under a blanket of nitrogen enriched in the $^{15}$N isotope to 99.5\% at a pressure of 850 torr to 1550\textdegree C, to produce a homogeneous molten solution. 
After 24 hours, the solution is slowly cooled at 4\textdegree C/h, to 1500\textdegree C, then at 50\textdegree C/h to 1300\textdegree C, and 200\textdegree C/h to room temperature. 
The \Nft solubility is decreased as the temperature is reduced, causing crystals to precipitate on the surface of the metal. 
The \Nft flakes were exfoliated from the metal with thermal release tape. 
The free-standing \Nft crystalline flakes were typically 10 to 20\,$\mu$m thick. 
To create boron vacancies, the \Nft flakes were neutron-irradiated. 
Samples were irradiated in The Ohio State University nuclear reactor, operated at a power 300\,kW for 1 hour, corresponding to neutron fluences of $1.4\times10^{16}$ neutron/cm$^{2}$. Prior to the cross-relaxometry experiments, micrometer-scale flakes were exfoliated from irradiated macroscopic crystals using standard mechanical exfoliation. This process yielded flakes with thicknesses of approximately 50~nm, used for the measurements shown in Fig.~2, and approximately 15~nm, used for the data presented in Fig.~S26.


Photoluminescence (PL) and Raman spectroscopy were performed on hBN$\text{nat}$ using a commercial InVia Qontor confocal system (Renishaw) to characterize defect implantation. 
The sample was excited using a 532\,nm laser delivered through a $100\times$ Leica objective (NA = 0.85). 
Scattered signals were collected by the same objective, spectrally filtered using ultra-narrow notch filters, and subsequently dispersed onto diffraction gratings with groove densities of 1800\,lines\,mm$^{-1}$ (Raman measurements) and 300\,lines\,mm$^{-1}$ (PL measurements).
The PL spectrum of the \Nft samples was not measured but is expected to have a similar spectrum reported in previous studies
\cite{Gong_Zu_2024}.


\subsection*{\label{ssec:Experimental_Setup}Experimental Setup}

NV measurements were performed with a commercial Qnami ProteusQ\texttrademark{} room temperature scanning NV microscope. 
A cantilever with a diamond nanopillar containing a single NV center was brought near (${<}5\,\text{nm}$) to the hBN surface. 
The NV center was excited using a 520\,nm laser ($\sim$20\,$\mu$W), and the photoluminescence was collected via a confocal optical setup. 
Microwaves ($\sim$0.1\,W input) were supplied by a nearby shorted coaxial cable brought within $\sim$15\,$\mu$m of the NV, using a Qnami MicrowaveQ signal generator or a Keysight M8195A Arbitrary Waveform Generator.


\subsection*{\label{ssec:Sensing_Protocols}Sensing Protocols}

Continuous-wave (CW) optically detected magnetic resonance (ODMR) measurements of the NV center and the \vbm ensemble, as shown in Figure~\ref{fig:1}c and \ref{fig:3}, were performed under continuous laser excitation and microwave drive. 
A static magnetic field was applied using a neodymium ring magnet mounted on a three-axis translation stage, while the microwave frequency was swept. To suppress fluorescence from the \vbm ensemble, a 750\,nm short-pass filter was used during NV ODMR measurements.

Pulsed NV $T_1$ relaxometry measurements were conducted using a 3\,$\mu$s laser pulse to initialize the NV spin into the $m_s = 0$ state. 
The spin was then allowed to relax toward thermal equilibrium during a dark interval $\tau$, during which microwave excitation was applied. 
Relaxation was detected as a reduction in photoluminescence (PL) during a subsequent readout sequence, preceding the next initialization pulse. 
At each readout pulse sequence, a reference PL signal, PL(0), is recorded after re-polarizing the spin into the $m_s = 0$ state. 
By varying $\tau$, exponential decay curves $\Delta \mathrm{PL}(\tau) = \mathrm{PL}(\tau)/\mathrm{PL}(0) \propto \exp(-\tau/T_1)$ such as those shown in Fig.\,\ref{fig:1}d were obtained and fit, from which the $T_1$ value was extracted. 

To reduce acquisition time, iso-$T_1$ relaxometry measurements were performed by fixing time $\tau$. 
In this case, the relative spin relaxation signal was estimated as $\Delta \mathrm{PL} = \frac{\mathrm{PL}(\tau)}{\mathrm{PL}(0)}$, providing a rapid proxy for magnetic noise mapping with reduced temporal overhead.

\processdelayedfloats


\newcommand{\acknowtext}{%
The scanning NV microscopy, hBN$_\text{nat}$ synthesis, and nanofabrication were supported by the Center for Nanophase Materials Sciences, (CNMS), which is a US Department of Energy, Office of Science User Facility at Oak Ridge National Laboratory. 
Spin relaxation measurements were supported by the Laboratory Directed Research and Development Program of Oak Ridge National Laboratory, managed by UT-Battelle, LLC, for the U.S.~Department of Energy. 
The RF controls were supported by the U.S.~Department of Energy, Office of Science, Basic Energy Sciences, Materials Sciences and Engineering Division. 
Support for \Nft crystal growth was provided by the Office of Naval Research, award number N00014-22-1-2582. 
Neutron irradiation of the \Nft crystals was supported by the U.S.~Department of Energy, Office of Nuclear Energy under DOE Idaho Operations Office Contract DE-AC07-051D13417 as part of a Nuclear Science User Facilities experiment.
This manuscript has been authored by UT-Battelle, LLC, under contract DE-AC05-00OR22725 with the US Department of Energy (DOE). 
The US government retains and the publisher, by accepting the article for publication, acknowledges that the US government retains a nonexclusive, paid-up, irrevocable, worldwide license to publish or reproduce the published form of this manuscript, or allow others to do so, for US government purposes. 
DOE will provide public access to these results of federally sponsored research in accordance with the DOE Public Access Plan (http://energy.gov/downloads/doe-public-access-plan).
}

\newcommand{\contritext}{%
H.Z.~conceived the project and performed the NV relaxometry experiments.
A.L.M.~and H.Z.~analyzed the data, and co-wrote the manuscript with input from all other authors.
A.L.M., Y.W., R.G.~and G.H.~developed the theoretical model, with additional input from L.L., C.Z., and J.T.D. 
R.G., G.H.~and C.Z.~performed the Monte Carlo simulation and quantified the defect density.
I.V.V.~synthesized the hBN$_\text{nat}$ samples and performed optical spectroscopy.
T.P.~and J.H.E.~synthesized the \Nft samples and organized the neutron irradiation.
S.R.~conducted the helium ion implantation.
S.G.~fabricated the coplanar waveguide.
Y.C.W.~and B.J.L.~assisted with the microwave delivery setup.
A.-P.L., S.J., and B.J.L.~provided technical support for the scanning NV microscope measurements.
}

\newcommand{\comptext}{%
The authors declare no competing interests.
}

\newcommand{\correspdtext}{%
%
Correspondence and requests for materials should be addressed to Huan Zhao.
}

\section*{Data availability}

The raw data of the main text figures are available in the Zenodo database \cite{melendez_2026_18520527}. Further data are available from the corresponding author upon request.

\clearpage 

\bibliographystyle{naturemag}
\bibliography{Bibliography_CR}

@article{Chen_Sun_2023,
title = {Quantum enhanced radio detection and ranging with solid spins},
journal = {Nat. Comm.},
volume = {14},
number = {1288},
pages = {},
year = {2023},
issn = {},
doi = {https://doi.org/10.1038/s41467-023-36929-8},
url = {https://www.nature.com/articles/s41467-023-36929-8},
author = {Xiang-Dong Chen and En-Hui Wang and Long-Kun Shan and Shao-Chun Zhang and Ce Feng and Yu Zheng and Yang Dong and Guang-Can Guo and Fang-Wen Sun},
}

@article{Yao_Lukin_2012,
  title={Scalable architecture for a room temperature solid-state quantum information processor},
  author={Yao, Norman Y and Jiang, Liang and Gorshkov, Alexey V and Maurer, Peter C and Giedke, Geza and Cirac, J Ignacio and Lukin, Mikhail D},
  journal={Nature communications},
  volume={3},
  number={1},
  pages={800},
  year={2012},
  publisher={Nature Publishing Group UK London}
}

@article{Alsid_Danielle_2023,
  title = {Solid-State Microwave Magnetometer with Picotesla-Level Sensitivity},
  author = {Alsid, Scott T. and Schloss, Jennifer M. and Steinecker, Matthew H. and Barry, John F. and Maccabe, Andrew C. and Wang, Guoqing and Cappellaro, Paola and Braje, Danielle A.},
  journal = {Phys. Rev. Appl.},
  volume = {19},
  issue = {5},
  pages = {054095},
  numpages = {8},
  year = {2023},
  month = {May},
  publisher = {American Physical Society},
  doi = {10.1103/PhysRevApplied.19.054095},
  url = {https://link.aps.org/doi/10.1103/PhysRevApplied.19.054095}
}

@article{Qiu_Yacoby_2022,
title = {Nanoscale electric field imaging with an ambient scanning quantum sensor microscope},
journal = {npj Quantum Inf.},
volume = {8},
number = {107},
pages = {},
year = {2022},
issn = {},
doi = {https://doi.org/10.1038/s41534-022-00622-3},
url = {https://www.nature.com/articles/s41534-022-00622-3},
author = {Ziwei Qiu and Assaf Hamo and Uri Vool and Tony X. Zhou and Amir Yacoby},
}

@article{Bian_Jiang_2021,
title = {Nanoscale electric-field imaging based on a quantum sensor and its charge-state control under ambient condition},
journal = {Nat. Comm.},
volume = {12},
number = {2457},
pages = {},
year = {2021},
issn = {},
doi = {https://doi.org/10.1038/s41467-021-22709-9},
url = {https://www.nature.com/articles/s41467-021-22709-9},
author = {Ke Bian and Wentian Zheng and Xianzhe Zeng and Xiakun Chen and Rainer Stöhr and Andrej Denisenko and Sen Yang and Jörg Wrachtrup and Ying Jiang},
}

@article{Liu_Li_2023,
author = {Liu, Gang-Qin and Liu, Ren-Bao and Li, Quan},
title = {Nanothermometry with Enhanced Sensitivity and Enlarged Working Range Using Diamond Sensors},
journal = {Acc. Chem. Res.},
volume = {56},
number = {2},
pages = {95-105},
year = {2023},
doi = {10.1021/acs.accounts.2c00576},
    note ={PMID: 36594628},
URL = {https://doi.org/10.1021/acs.accounts.2c00576},
eprint = { 
    https://doi.org/10.1021/acs.accounts.2c00576
}}

@article{Gottscholl_Dyakonov_2021,
title = {Spin defects in hBN as promising temperature, pressure and magnetic field quantum sensors},
journal = {Nat. Comm.},
volume = {12},
number = {4480},
pages = {},
year = {2021},
issn = {},
doi = {https://doi.org/10.1038/s41467-021-24725-1},
url = {https://www.nature.com/articles/s41467-021-24725-1},
author = {Andreas Gottscholl and Matthias Diez and Victor Soltamov and Christian Kasper and Dominik Krauße and Andreas Sperlich and Mehran Kianinia and Carlo Bradac and Igor Aharonovich and Vladimir Dyakonov},
}

@article{
Fujiwara_2020,
author = {Masazumi Fujiwara  and Simo Sun  and Alexander Dohms  and Yushi Nishimura  and Ken Suto  and Yuka Takezawa  and Keisuke Oshimi  and Li Zhao  and Nikola Sadzak  and Yumi Umehara  and Yoshio Teki  and Naoki Komatsu  and Oliver Benson  and Yutaka Shikano  and Eriko Kage-Nakadai },
title = {Real-time nanodiamond thermometry probing in vivo thermogenic responses},
journal = {Sci. Adv.},
volume = {6},
number = {37},
pages = {eaba9636},
year = {2020},
doi = {10.1126/sciadv.aba9636},
URL = {https://www.science.org/doi/abs/10.1126/sciadv.aba9636},
eprint = {https://www.science.org/doi/pdf/10.1126/sciadv.aba9636}}

@article{Tzeng_Chang_2015,
author = {Tzeng, Yan-Kai and Tsai, Pei-Chang and Liu, Hsiou-Yuan and Chen, Oliver Y. and Hsu, Hsiang and Yee, Fu-Goul and Chang, Ming-Shien and Chang, Huan-Cheng},
title = {Time-Resolved Luminescence Nanothermometry with Nitrogen-Vacancy Centers in Nanodiamonds},
journal = {Nano Lett.},
volume = {15},
number = {6},
pages = {3945-3952},
year = {2015},
doi = {10.1021/acs.nanolett.5b00836},
    note ={PMID: 25951304},
URL = {https://doi.org/10.1021/acs.nanolett.5b00836},
eprint = {https://doi.org/10.1021/acs.nanolett.5b00836
}}

@article{Degen_Cappellaro_2017,
  title = {Quantum sensing},
  author = {Degen, C. L. and Reinhard, F. and Cappellaro, P.},
  journal = {Rev. Mod. Phys.},
  volume = {89},
  issue = {3},
  pages = {035002},
  numpages = {39},
  year = {2017},
  month = {Jul},
  publisher = {American Physical Society},
  doi = {10.1103/RevModPhys.89.035002},
  url = {https://link.aps.org/doi/10.1103/RevModPhys.89.035002}
}

@article{Aslam_Park_2023,
  title = {Quantum sensors for biomedical applications},
  author = {Nabeel Aslam and Hengyun Zhou and Elana K. Urbach and Matthew J. Turner and Ronald L. Walsworth and Mikhail D. Lukin and Hongkun Park},
  journal = {Nat. Rev. Phys.},
  volume = {5},
  issue = {},
  pages = {157-169},
  numpages = {},
  year = {2023},
  month = {Feb},
  publisher = {},
  doi = {10.1038/s42254-023-00558-3},
  url = {https://www.nature.com/articles/s42254-023-00558-3}
}

@article{Maletinsky_Jacques_2014,
    Author = {Rondin, L. and Tetienne, J-P and Hingant, T. and Roch, J-F and
       Maletinsky, P. and Jacques, V.},
    Title = {Magnetometry with nitrogen-vacancy defects in diamond},
    Journal = {Rep. Prog. Phys.},
    Year = {2014},
    Volume = {77},
    Number = {5},
    Month = {MAY},
    DOI = {10.1088/0034-4885/77/5/056503},
    Article-Number = {056503},
    ISSN = {0034-4885},
    EISSN = {1361-6633},
    ResearcherID-Numbers = {Tetienne, Jean-Philippe/Q-4171-2019
       Maletinsky, Patrick/L-1851-2015
       Tetienne, Jean-Philippe/H-4896-2014
       Jacques, Vincent/D-3881-2014
       },
    ORCID-Numbers = {Tetienne, Jean-Philippe/0000-0001-5796-2508
       Maletinsky, Patrick/0000-0003-1699-388X
       Tetienne, Jean-Philippe/0000-0001-5796-2508
       Rondin, Loic/0000-0002-4833-2886},
    Unique-ID = {WOS:000337351800003},
}

@article{Yacoby_Casola_vanderSar_2018,
	title = {Probing condensed matter physics with magnetometry based on nitrogen-vacancy centres in diamond},
	volume = {3},
	issn = {},
	url = {https://www.nature.com/articles/natrevmats201788},
	doi = {},
	number = {},
	urldate = {},
	journal = {Nat. Rev. Mater.},
	author = {Casola, F. and van der Sar, T. and Yacoby, A.},
	month = {},
	year = {2018},
	pages = {17088}
}

@article{Jelezko_Wrachtrup_2006,
Author = {Jelezko, F. and Wrachtrup, J.},
Title = {Single defect centres in diamond: A review},
Journal = {Phys. Status Solidi A},
Year = {2006},
Volume = {203},
Number = {13, SI},
Pages = {3207-3225},
Month = {OCT},
DOI = {10.1002/pssa.200671403},
ISSN = {1862-6300},
EISSN = {1862-6319},
Unique-ID = {WOS:000241640200002},
}

@article{Doherty_2013,
title = {The nitrogen-vacancy colour centre in diamond},
journal = {Phys. Rep.},
volume = {528},
number = {1},
pages = {1-45},
year = {2013},
issn = {0370-1573},
doi = {https://doi.org/10.1016/j.physrep.2013.02.001},
url = {https://www.sciencedirect.com/science/article/pii/S0370157313000562},
author = {Marcus W. Doherty and Neil B. Manson and Paul Delaney and Fedor Jelezko and Jörg Wrachtrup and Lloyd C.L. Hollenberg},
}

@article{Chen_Yan_2025,
url = {https://doi.org/10.1515/nanoph-2024-0569},
title = {Low-dimensional solid-state single-photon emitters},
author = {Jinli Chen and Chaohan Cui and Ben Lawrie and Yongzhou Xue and Saikat Guha and Matt Eichenfield and Huan Zhao and Xiaodong Yan},
journal = {Nanophotonics},
doi = {doi:10.1515/nanoph-2024-0569},
year = {2025},
lastchecked = {2025-04-08}
}

@article{Fang_Sun_2024,
title = {Quantum sensing with optically accessible spin defects in van der Waals layered materials},
journal = {Light Sci. Appl.},
volume = {13},
number = {303},
pages = {},
year = {2024},
issn = {},
doi = {https://doi.org/10.1038/s41377-024-01630-y},
url = {https://www.nature.com/articles/s41377-024-01630-y},
author = {Hong-Hua Fang and Xiao-Jie Wang and Xavier Marie and Hong-Bo Sun},
}

@article{Guo_Gu_2024,
title = {Quantum defects in two-dimensional van der Waals materials},
journal = {Fundam. Res.},
year = {2024},
issn = {2667-3258},
doi = {https://doi.org/10.1016/j.fmre.2024.01.019},
url = {https://www.sciencedirect.com/science/article/pii/S2667325824000426},
author = {Yang Guo and Jianmei Li and Ruifen Dou and Haitao Ye and Changzhi Gu}
}

@article{Gottscholl_Dyakonov_2020,
title = {Initialization and read-out of intrinsic spin defects in a van der Waals crystal at room temperature},
journal = {Nat. Mat.},
volume = {19},
number = {},
pages = {540–545},
year = {2020},
issn = {},
doi = {https://doi.org/10.1038/s41563-020-0619-6},
url = {https://www.nature.com/articles/s41563-020-0619-6},
author = {Andreas Gottscholl and Mehran Kianinia and Victor Soltamov and Sergei Orlinskii and Georgy Mamin and Carlo Bradac and Christian Kasper and Klaus Krambrock and Andreas Sperlich and Milos Toth and Igor Aharonovich and Vladimir Dyakonov},
}

@article{Segawa_2023,
title = {Nanoscale quantum sensing with Nitrogen-Vacancy centers in nanodiamonds – A magnetic resonance perspective},
journal = {Prog. Nucl. Magn. Reson. Spectrosc.},
volume = {134-135},
pages = {20-38},
year = {2023},
issn = {0079-6565},
doi = {https://doi.org/10.1016/j.pnmrs.2022.12.001},
url = {https://www.sciencedirect.com/science/article/pii/S0079656522000322},
author = {Takuya F. Segawa and Ryuji Igarashi}
}

@article{Lyu_Gao_2022,
author = {Lyu, Xiaodan and Tan, Qinghai and Wu, Lishu and Zhang, Chusheng and Zhang, Zhaowei and Mu, Zhao and Zú{\~n}iga-P{\'e}rez, Jesús and Cai, Hongbing and Gao, Weibo},
title = {Strain Quantum Sensing with Spin Defects in Hexagonal Boron Nitride},
journal = {Nano Lett.},
volume = {22},
number = {16},
pages = {6553-6559},
year = {2022},
doi = {10.1021/acs.nanolett.2c01722},
    note ={PMID: 35960708},
URL = {https://doi.org/10.1021/acs.nanolett.2c01722},
eprint = {https://doi.org/10.1021/acs.nanolett.2c01722}}

@article{Bajaj_2014,
title = {Optically detected cross-relaxation spectroscopy of electron spins in diamond},
journal = {Nat. Comm.},
volume = {5},
number = {4135},
pages = {},
year = {2014},
issn = {},
doi = {https://doi.org/10.1038/ncomms5135},
url = {https://www.nature.com/articles/ncomms5135},
author = {Hai-Jing Wang and Chang S. Shin and Scott J. Seltzer and Claudia E. Avalos and Alexander Pines and Vikram S. Bajaj},
}

@article{Liu_Zu_2025,
  title = {Temperature-dependent spin-phonon coupling of boron-vacancy centers in hexagonal boron nitride},
  author = {Liu, Zhongyuan and Gong, Ruotian and Huang, Benchen and Jin, Yu and Du, Xinyi and He, Guanghui and Janzen, Eli and Yang, Li and Henriksen, Erik A. and Edgar, James H. and Galli, Giulia and Zu, Chong},
  journal = {Phys. Rev. B},
  volume = {111},
  issue = {2},
  pages = {024108},
  numpages = {7},
  year = {2025},
  month = {Jan},
  publisher = {American Physical Society},
  doi = {10.1103/PhysRevB.111.024108},
  url = {https://link.aps.org/doi/10.1103/PhysRevB.111.024108}
}

@article{Gong_Zu_2024,
  title={Isotope engineering for spin defects in van der Waals materials},
  author={Gong, Ruotian and Du, Xinyi and Janzen, Eli and Liu, Vincent and Liu, Zhongyuan and He, Guanghui and Ye, Bingtian and Li, Tongcang and Yao, Norman Y and Edgar, James H and Henriksen, Erik A. and Zu, Chong},
  journal={Nat. Comm.},
  volume={15},
  number={1},
  pages={104},
  year={2024},
  publisher={Nature Publishing Group UK London}
}

@article{Wood_Hollenberg_2016,
title = {Wide-band nanoscale magnetic resonance spectroscopy using quantum relaxation of a single spin in diamond},
journal = {Phys. Rev. B},
volume = {94},
number = {155402},
pages = {},
year = {2016},
issn = {},
doi = {https://doi.org/10.1103/PhysRevB.94.155402},
url = {https://journals.aps.org/prb/abstract/10.1103/PhysRevB.94.155402},
author = {James D. A. Wood and David A. Broadway and Liam T. Hall and Alastair Stacey and David A. Simpson and Jean-Philippe Tetienne and Lloyd C. L. Hollenberg},
}

@article{Wood_Hollenberg_2017,
  title     = {Microwave-free nuclear magnetic resonance at molecular scales},
  author    = {Wood, James D. A. and Tetienne, Jean-Philippe and Broadway, David A. and Hall, Liam T. and Simpson, David A. and Stacey, Alastair and Hollenberg, Lloyd C. L.},
  journal   = {Nature Communications},
  volume    = {8},
  number    = {},
  pages     = {15950},
  year      = {2017},
  doi       = {10.1038/ncomms15950},
  url       = {https://www.nature.com/articles/ncomms15950}
}

@article{Epstein_Awschalom_2005,
title = {Anisotropic interactions of a single spin and dark-spin spectroscopy in diamond},
journal = {Nat. Phys.},
volume = {1},
number = {},
pages = {94-98},
year = {2005},
issn = {},
doi = {https://doi.org/10.1038/nphys141},
url = {https://www.nature.com/articles/nphys141},
author = {R. J. Epstein and F. M. Mendoza and Y. K. Kato and D. D. Awschalom},
}

@article{Sarkar_Gradecak_2024,
author = {Sarkar, Soumya and Xu, Yue and Mathew, Sinu and Lal, Manohar and Chung, Jing-Yang and Lee, Hae Yeon and Watanabe, Kenji and Taniguchi, Takashi and Venkatesan, Thirumalai and Gradečak, Silvija},
title = {Identifying Luminescent Boron Vacancies in h-BN Generated Using Controlled {He$^+$} Ion Irradiation},
journal = {Nano Lett.},
volume = {24},
number = {1},
pages = {43-50},
year = {2024},
doi = {https://doi.org/10.1021/acs.nanolett.3c03113},
    note ={PMID: 37930062},
URL = { 
    https://pubs.acs.org/doi/epdf/10.1021/acs.nanolett.3c03113
},
eprint = {
    https://doi.org/10.1021/acs.nanolett.3c03113
}
}

@article{Eckmann_Casiraghi_2012,
author = {Eckmann, Axel and Felten, Alexandre and Mishchenko, Artem and Britnell, Liam and Krupke, Ralph and Novoselov, Kostya S. and Casiraghi, Cinzia},
title = {Probing the Nature of Defects in Graphene by Raman Spectroscopy},
journal = {Nano Lett.},
volume = {12},
number = {8},
pages = {3925-3930},
year = {2012},
doi = {10.1021/nl300901a},
    note ={PMID: 22764888},
URL = {https://doi.org/10.1021/nl300901a
},
eprint = {https://doi.org/10.1021/nl300901a}}

@article{Zhang_2024,
title = {Insights into the role of defects on the Raman spectroscopy of carbon nanotube and biomass-derived carbon},
journal = {Carbon},
volume = {222},
pages = {118998},
year = {2024},
issn = {0008-6223},
doi = {https://doi.org/10.1016/j.carbon.2024.118998},
url = {https://www.sciencedirect.com/science/article/pii/S0008622324002173},
author = {Peng Zhang and Jingyuan Fan and Yuanqing Wang and Yuying Dang and Saskia Heumann and Yuxiao Ding}
}

@article{Dresselhaus_Saito_2010,
author = {Dresselhaus, M. S.  and Jorio, A.  and Souza Filho, A. G.  and Saito, R. },
title = {Defect characterization in graphene and carbon nanotubes using Raman spectroscopy},
journal = {Philos. Trans. R. Soc. A},
volume = {368},
number = {1932},
pages = {5355-5377},
year = {2010},
doi = {10.1098/rsta.2010.0213},
URL = {https://royalsocietypublishing.org/doi/abs/10.1098/rsta.2010.0213},
eprint = {https://royalsocietypublishing.org/doi/pdf/10.1098/rsta.2010.0213}
}

@article{Lee_Jeong_2022,
title = {Investigating heterogeneous defects in single-crystalline {WS$_2$} via tip-enhanced Raman spectroscopy},
journal = {npj {2D} Mater. Appl.},
volume = {6},
number = {67},
pages = {},
year = {2022},
issn = {},
doi = {https://doi.org/10.1038/s41699-022-00334-4},
url = {https://www.nature.com/articles/s41699-022-00334-4},
author = {Chanwoo Lee and Byeong Geun Jeong and Sung Hyuk Kim and Dong Hyeon Kim and Seok Joon Yun and Wooseon Choi and Sung-Jin An and Dongki Lee and Young-Min Kim and Ki Kang Kim and Seung Mi Lee and Mun Seok Jeong},
}

@article{Gong_Zu_2023,
title = {Coherent dynamics of strongly interacting electronic spin defects in hexagonal boron nitride},
journal = {Nat. Comm.},
volume = {14},
number = {3299},
pages = {},
year = {2023},
issn = {},
doi = {https://doi.org/10.1038/s41467-023-39115-y},
url = {https://www.nature.com/articles/s41467-023-39115-y},
author = {Ruotian Gong and Guanghui He and Xingyu Gao and Peng Ju and Zhongyuan Liu and Bingtian Ye and Erik A. Henriksen and Tongcang Li and Chong Zu},
}

@article{Clua-Provost_Jacques_2024,
  title = {Spin-dependent photodynamics of boron-vacancy centers in hexagonal boron nitride},
  author = {Clua-Provost, T. and Mu, Z. and Durand, A. and Schrader, C. and Happacher, J. and Bocquel, J. and Maletinsky, P. and Frauni\'e, J. and Marie, X. and Robert, C. and Seine, G. and Janzen, E. and Edgar, J. H. and Gil, B. and Cassabois, G. and Jacques, V.},
  journal = {Phys. Rev. B},
  volume = {110},
  issue = {1},
  pages = {014104},
  numpages = {11},
  year = {2024},
  month = {Jul},
  publisher = {American Physical Society},
  doi = {10.1103/PhysRevB.110.014104},
  url = {https://link.aps.org/doi/10.1103/PhysRevB.110.014104}
}

@article{Clua-Provost_Jacques_2023,
  title={Isotopic control of the boron-vacancy spin defect in hexagonal boron nitride},
  author={Clua-Provost, T and Durand, A and Mu, Z and Rastoin, T and Frauni{\'e}, J and Janzen, E and Schutte, H and Edgar, JH and Seine, G and Claverie, A and others},
  journal={Physical Review Letters},
  volume={131},
  number={12},
  pages={126901},
  year={2023},
  publisher={APS}
}

@Article{Krecmarova_Sanchez_2019,
AUTHOR = {Krečmarová, Marie and Andres-Penares, Daniel and Fekete, Ladislav and Ashcheulov, Petr and Molina-Sánchez, Alejandro and Canet-Albiach, Rodolfo and Gregora, Ivan and Mortet, Vincent and Martínez-Pastor, Juan P. and Sánchez-Royo, Juan F.},
TITLE = {Optical Contrast and Raman Spectroscopy Techniques Applied to Few-Layer {2D} Hexagonal Boron Nitride},
JOURNAL = {Nanomaterials},
VOLUME = {9},
YEAR = {2019},
NUMBER = {7},
ARTICLE-NUMBER = {1047},
URL = {https://www.mdpi.com/2079-4991/9/7/1047},
PubMedID = {31336572},
ISSN = {2079-4991},
DOI = {10.3390/nano9071047}
}

@article{Schirhagl_Degen_2014,
   author = "Schirhagl, Romana and Chang, Kevin and Loretz, Michael and Degen, Christian L.",
   title = "Nitrogen-Vacancy Centers in Diamond: Nanoscale Sensors for Physics and Biology", 
   journal= "Annu. Rev. Phys. Chem.",
   year = "2014",
   volume = "65",
   number = "Volume 65, 2014",
   pages = "83-105",
   doi = "https://doi.org/10.1146/annurev-physchem-040513-103659",
   url = "https://www.annualreviews.org/content/journals/10.1146/annurev-physchem-040513-103659",
   publisher = "Annual Reviews",
   issn = "1545-1593",
   type = "Journal Article"
  }

@article{Du_Wrachtrup_2024,
  title = {Single-molecule scale magnetic resonance spectroscopy using quantum diamond sensors},
  author = {Du, Jiangfeng and Shi, Fazhan and Kong, Xi and Jelezko, Fedor and Wrachtrup, J\"org},
  journal = {Rev. Mod. Phys.},
  volume = {96},
  issue = {2},
  pages = {025001},
  numpages = {62},
  year = {2024},
  month = {May},
  publisher = {American Physical Society},
  doi = {10.1103/RevModPhys.96.025001},
  url = {https://link.aps.org/doi/10.1103/RevModPhys.96.025001}
}

@article{Linderalv_Erhart_2021,
  title = {Vibrational signatures for the identification of single-photon emitters in hexagonal boron nitride},
  author = {Linder\"alv, Christopher and Wieczorek, Witlef and Erhart, Paul},
  journal = {Phys. Rev. B},
  volume = {103},
  issue = {11},
  pages = {115421},
  numpages = {13},
  year = {2021},
  month = {Mar},
  publisher = {American Physical Society},
  doi = {10.1103/PhysRevB.103.115421},
  url = {https://link.aps.org/doi/10.1103/PhysRevB.103.115421}
}

@article{He_Zu_2023,
  title={Quasi-{F}loquet prethermalization in a disordered dipolar spin ensemble in diamond},
  author={He, Guanghui and Ye, Bingtian and Gong, Ruotian and Liu, Zhongyuan and Murch, Kater W and Yao, Norman Y and Zu, Chong},
  journal={Physical Review Letters},
  volume={131},
  number={13},
  pages={130401},
  year={2023},
  publisher={APS}
}

@article{Liu_Edgar_2018,
  title={Single crystal growth of millimeter-sized monoisotopic hexagonal boron nitride},
  author={Liu, Song and He, Rui and Xue, Lianjie and Li, Jiahan and Liu, Bin and Edgar, James H},
  journal={Chemistry of materials},
  volume={30},
  number={18},
  pages={6222--6225},
  year={2018},
  publisher={ACS Publications}
}

@article{Janzen_Edgar_2024,
  title={Boron and nitrogen isotope effects on hexagonal boron nitride properties},
  author={Janzen, Eli and Schutte, Hannah and Plo, Juliette and Rousseau, Adrien and Michel, Thierry and Desrat, Wilfried and Valvin, Pierre and Jacques, Vincent and Cassabois, Guillaume and Gil, Bernard and others},
  journal={Advanced Materials},
  volume={36},
  number={2},
  pages={2306033},
  year={2024},
  publisher={Wiley Online Library}
}

@article{grinolds2013nanoscale,
  title={Nanoscale magnetic imaging of a single electron spin under ambient conditions},
  author={Grinolds, Michael Sean and Hong, Sungkun and Maletinsky, Patrick and Luan, Lan and Lukin, Mikhail D and Walsworth, Ronald Lee and Yacoby, Amir},
  journal={Nature Physics},
  volume={9},
  number={4},
  pages={215--219},
  year={2013},
  publisher={Nature Publishing Group UK London}
}

@article{sushkov2014magnetic,
  title={Magnetic resonance detection of individual proton spins using quantum reporters},
  author={Sushkov, Alexander O and Lovchinsky, Igor and Chisholm, Nicholas and Walsworth, Ronald Lee and Park, Hongkun and Lukin, Mikhail D},
  journal={Physical review letters},
  volume={113},
  number={19},
  pages={197601},
  year={2014},
  publisher={APS}
}

@article{shi2015single,
  title={Single-protein spin resonance spectroscopy under ambient conditions},
  author={Shi, Fazhan and Zhang, Qi and Wang, Pengfei and Sun, Hongbin and Wang, Jiarong and Rong, Xing and Chen, Ming and Ju, Chenyong and Reinhard, Friedemann and Chen, Hongwei and others},
  journal={Science},
  volume={347},
  number={6226},
  pages={1135--1138},
  year={2015},
  publisher={American Association for the Advancement of Science}
}

@article{wang2019nanoscale,
  title={Nanoscale magnetic imaging of ferritins in a single cell},
  author={Wang, Pengfei and Chen, Sanyou and Guo, Maosen and Peng, Shijie and Wang, Mengqi and Chen, Ming and Ma, Wenchao and Zhang, Rui and Su, Jihu and Rong, Xing and others},
  journal={Science advances},
  volume={5},
  number={4},
  pages={eaau8038},
  year={2019},
  publisher={American Association for the Advancement of Science}
}

@article{Rugar_Awschalom_2015,
title = {Proton magnetic resonance imaging using a nitrogen-vacancy spin sensor},
journal = {Nat. Nanotech.},
volume = {10},
number = {},
pages = {120-124},
year = {2015},
issn = {},
doi = {https://doi.org/10.1038/nnano.2016.63},
url = {https://www.nature.com/articles/nnano.2014.288},
author = {D. Rugar and H. J. Mamin and M. H. Sherwood and M. Kim and C. T. Rettner and K. Ohno and D. D. Awschalom},
}

@article{Vlassiouk_2025,
doi = {10.48550/arXiv.2503.18894},
url = {https://arxiv.org/abs/2503.18894},
year = {2025},
month = {mar},
publisher = {},
volume = {},
number = {},
pages = {},
author = {Ivan V. Vlassiouk and Yueh-Chun Wu and Alexander Puretzky and Liangbo Liang and John Lasseter and Bogdan Dryzhakov and Ian Gallagher and Sujoy Ghosh and Nickolay Lavrik and Ondrej Dyck and Andrew R. Lupini and Marti Checa and Liam Collins and Huan Zhao and Farzana Likhi and Kai Xiao and Ilia Ivanov and David Glasgow and Alexander Tselev and Benjamin Lawrie and Sergei Smirnov and Steven Randolph},
title = {Defect Engineering in Large-Scale CVD-Grown Hexagonal Boron Nitride: Formation, Spectroscopy, and Spin Relaxation Dynamics},
journal = {arXiv}
}

@article{Wu_Lawrie_2025,
author = {Wu, Yueh-Chun and Halász, Gábor B. and Damron, Joshua T. and Gai, Zheng and Zhao, Huan and Sun, Yuxin and Dahmen, Karin A. and Sohn, Changhee and Carlson, Erica W. and Hua, Chengyun and Lin, Shan and Song, Jeongkeun and Lee, Ho Nyung and Lawrie, Benjamin J.},
title = {Nanoscale Magnetic Ordering Dynamics in a High Curie Temperature Ferromagnet},
journal = {Nano Letters},
volume = {25},
number = {4},
pages = {1473-1479},
year = {2025},
doi = {10.1021/acs.nanolett.4c05401},
    note ={PMID: 39804711},

URL = { https://doi.org/10.1021/acs.nanolett.4c05401
},
eprint = { https://doi.org/10.1021/acs.nanolett.4c05401}
}

@article{Barry_Walsworth_2020,
  title = {Sensitivity optimization for NV-diamond magnetometry},
  author = {Barry, John F. and Schloss, Jennifer M. and Bauch, Erik and Turner, Matthew J. and Hart, Connor A. and Pham, Linh M. and Walsworth, Ronald L.},
  journal = {Rev. Mod. Phys.},
  volume = {92},
  issue = {1},
  pages = {015004},
  numpages = {68},
  year = {2020},
  month = {Mar},
  publisher = {American Physical Society},
  doi = {10.1103/RevModPhys.92.015004},
  url = {https://link.aps.org/doi/10.1103/RevModPhys.92.015004}
}

@article{Scholten2024,
   author = {Sam C. Scholten and Priya Singh and Alexander J. Healey and Islay O. Robertson and Galya Haim and Cheng Tan and David A. Broadway and Lan Wang and Hiroshi Abe and Takeshi Ohshima and Mehran Kianinia and Philipp Reineck and Igor Aharonovich and Jean Philippe Tetienne},
   doi = {10.1038/s41467-024-51129-8},
   issn = {20411723},
   issue = {1},
   journal = {Nat. Commun.},
   month = {12},
   pmid = {39112477},
   publisher = {Nature Research},
   title = {Multi-species optically addressable spin defects in a van der Waals material},
   volume = {15},
   pages   = {6727},
   year = {2024}
}

@article{yuan2020charge,
  title={Charge state dynamics and optically detected electron spin resonance contrast of shallow nitrogen-vacancy centers in diamond},
  author={Yuan, Zhiyang and Fitzpatrick, Mattias and Rodgers, Lila VH and Sangtawesin, Sorawis and Srinivasan, Srikanth and De Leon, Nathalie P},
  journal={Physical Review Research},
  volume={2},
  number={3},
  pages={033263},
  year={2020},
  publisher={APS}
}

@article{fraunie2025charge,
  title={Charge state tuning of spin defects in hexagonal boron nitride},
  author={Frauni{\'e}, Jules and Clua-Provost, Tristan and Roux, S{\'e}bastien and Mu, Zhao and Delpoux, Adrien and Seine, Gr{\'e}gory and Lagarde, Delphine and Watanabe, Kenji and Taniguchi, Takashi and Marie, Xavier and others},
  journal={Nano Letters},
  volume={25},
  number={14},
  pages={5836--5842},
  year={2025},
  publisher={ACS Publications}
}

@article{sun2025room,
  title={Room-temperature hybrid 2d-3d quantum spin system for enhanced magnetic sensing and many-body dynamics},
  author={Sun, Haoyu and Yu, Pei and Zhou, Xu and Ye, Xiangyu and Wang, Mengqi and Liu, Zhaoxin and Guo, Yuhang and Liu, Wenzhao and Huang, You and Wang, Pengfei and others},
  journal={npj Quantum Information},
  year={2025},
  publisher={Nature Publishing Group UK London}
}

@article{Gong_2026_review,
doi = {10.48550/arXiv.2602.01521},
url = {https://arxiv.org/abs/2602.01521},
year = {2026},
month = {feb},
publisher = {},
volume = {},
number = {},
pages = {},
author = {Ruotian Gong and Alex L. Melendez and Guanghui He and Zhongyuan Liu and Chong Zu and Huan Zhao},
title = {Spin Relaxometry with Solid-State Defects: Theory, Platforms, and Applications},
journal = {arXiv}
}

@dataset{melendez_2026_18520527,
  author       = {Melendez, Alex Lee and
                  Gong, Ruotian and
                  He, Guanghui and
                  Wang, Yan and
                  Wu, Yueh-Chun and
                  Poirier, Thomas and
                  Randolph, Steven and
                  Ghosh, Sujoy and
                  Liang, Liangbo and
                  Jesse, Stephen and
                  Li, An-Ping and
                  Damron, Joshua and
                  Lawrie, Ben and
                  Edgar, James and
                  Vlassiouk, Ivan and
                  Zu, Chong and
                  Zhao, Huan},
  title        = {Data for Probing Boron Vacancy Defects in hBN via
                   Single Spin Relaxometry
                  },
  month        = feb,
  year         = 2026,
  publisher    = {Zenodo},
  doi          = {10.5281/zenodo.18520527},
  url          = {https://doi.org/10.5281/zenodo.18520527},
}


\paragraph*{Acknowledgments:}\acknowtext
\paragraph*{Author contributions:}\contritext
\paragraph*{Competing interests:}\comptext

\section*{Figure Legends}

\textbf{Fig.~1: NV $T_1$ relaxometry at CR condition with \vbm in hBN$_\text{nat}$:}
(a) A scanning NV microscope containing a single NV center is brought near a 90-300\,nm thick hBN$_\text{nat}$ sample that has been irradiated with He ions to create an ensemble of \vbm centers. 
The NV center's photoluminescence (PL) is collected using a confocal optical setup.
A microwave field $B_\text{MW}$ is supplied to both the NV and the nearest \vbm centers by an antenna positioned near the cantilever tip.
When the microwave frequency matches the spin resonance frequency of the NV or \vbm centers, transitions between the ground state sublevels $m_s=0\leftrightarrow\pm1$ states are driven coherently.
(b) Calculated spin resonance dispersions of the NV and \vbm centers with the magnetic field oriented 29\textdegree\ from the surface normal. 
Cross-relaxation occurs near 127\,G bias, where the NV $m_s = 0 \leftrightarrow +1$ transition overlaps with the \vbm $m_s = 0 \leftrightarrow -1$ transition, highlighted by the green dot.
(c) CW-ODMR spectra of the NV center before (green curve) and after (purple curve) engaging with the hBN sample, and of the \vbm centers in hBN (red curve), measured under the cross-relaxation condition shown in (b). 
All spectra are fitted using Lorentzian functions. The left and right vertical axes correspond to the NV and \vbm data, respectively. Only the $m_s = 0 \leftrightarrow +1$ transition of the NV is shown.
The arrows denote via their colors which vertical axis each trace is plotted against.
(d) NV spin $T_1$ measurement in non-CR condition (purple curve, fitted $T_1 = 1.34 \pm 0.19$\,ms) and CR condition (red curve, $T_1 = 406 \pm 34\,\mu$s), both measured after engaging the hBN$_\text{nat}$ sample. Curves are single exponential fits $
\propto\exp(-t/T_1)$.

\textbf{Fig.~2: $T_1$-MR detection of hyperfine structure:}
(a) CW-ODMR spectrum of \vbm centers in \Nft showing four hyperfine-split lines, fitted with Lorentzian functions (red). 
A magnetic field of 123\,G is applied out-of-plane. 
(b) Extracted ODMR center frequencies of the NV center (red) and \vbm centers (black) as a function of magnetic field, with solid lines representing fitted curves. 
The hyperfine line positions (dashed) are calculated using the average $A_{zz}$, revealing four expected cross-relaxation conditions corresponding to the crossover points between the NV and \vbm transitions. 
(c) NV single-$\tau$ $T_1$-MR detection of \vbm hyperfine structure by sweeping the magnetic field through the four CR conditions. 
The magnetic field is converted to frequency detuning between the NV and \vbm transitions and plotted in units of $A_{zz}$ in \Nft (67\,MHz).
The data are fitted using a sum of four Lorentzian functions, where dips 1 and 3 share one set of parameters, and dips 2 and 4 share another (fit shown in red). 
Here, the free evolution time $\tau$ is chosen to be 2\,ms, given that the tip's non-CR $T_1$ measured immediately before this $T_1$-MR measurement is $2.09 \pm 0.25$\,ms (see Supplementary Figure S23a). 
(d) NV single-$\tau$ $T_1$-MR measurement of hBN$_\text{nat}$ under varying magnetic field. 
The field axis is converted to frequency detuning and plotted in units of $A_{zz}$ in hBN$_\text{nat}$ (44\,MHz).
Data are fitted using a Lorentzian function (red curve). The $\tau$ in this case is 0.7\,ms, and the tip's $T_1$ measured at the non-CR condition is $1.21 \pm 0.32$\,ms (see Supplementary Figure S23b).
Error bars of (c) and (d) represent normalized shot noise level.
(e) A short-$T_1$ NV tip's spin $T_1$ relaxation curve before engaging the sample, showing $T_1$ of $191 \pm 7$\,$\mu$s.
(f) NV $T_1$ values measured as a function of magnetic field (blue symbols with error bars), showing four distinct cross-relaxation dips. The curve is fitted with a sum of Lorentzian functions and plotted in units of $A_{zz}$ in \Nft (67\,MHz). Error bars represent $1\sigma$ uncertainties of fitted $T_1$ values.

\textbf{Fig.~3: Nanoscale imaging of \vbm density in hBN$_\text{nat}$ using NV-\vbm cross relaxometry:}
(a) Optical microscope image of hBN$_\text{nat}$ sample on gold surface, with different colors corresponding to variations in the hBN thickness. 
The light-colored area in the center has thickness $\sim$90\,nm compared to the surrounding pink/purple areas with thickness $\sim$190\,nm.
Here a portion of the right edge of the light-colored area is highlighted in a blue polygon which was measured by the scanning NV in panel (f). 
(b) Raman spectroscopy map showing the ratio of the $E'$ peak to the $E_{\text{2g}}$ peak across the sample area displayed in panel (a). 
(c) SRIM simulations of the ion energy losses (normalized per ion) of 30\,keV He$^+$ (10,000 ions simulated) for 90\,nm and (d) 250\,nm hBN on gold, suggesting thickness-dependent non-uniform defect density distribution.
(e) Monte Carlo simulation of the cross-relaxometry contribution to the NV relaxation rate $\Gamma_1^\text{CR}$ as a function of the \vbm density and three different NV-to-sample distances. 
Horizontal gray bar shows values of $\Gamma_1^\text{CR}$ associated with range of the color scale in (f).
(f) \vbm density map, obtained as a single-$\tau$ $T_1$ spatial scan over a free-evolution time $\tau=250 \,\mu$s which was converted to \vbm density via the Monte Carlo simulation in panel (e). 
The size of each pixel is 100\,nm. The measurement was conducted at the CR condition.
Inset: The profile of \vbm density as a function of position, over the line indicated by the white dashed line in (f). The data is fit to the edge-spread function (red) yielding a step width of 46\,nm.

\end{document}